\documentstyle[11pt,epsfig,amssymb]{article}
\title{\bf Modified gravity and Space-Time-Matter theory}
\author{F. Darabi\thanks{email: f.darabi@azaruniv.edu}
\\ {\small Department of Physics, Azarbaijan
University of Tarbiat Moallem, Tabriz 53741-161, Iran}}
\begin{document}
\maketitle

\begin{abstract}
The correspondence between $f(R)$ theories of gravity and model
theories explaining induced dark energy in a $5D$ Ricci-flat
universe, known as the Space-Time-Matter theory (STM), is studied.
It is shown that such correspondence may be used to interpret the
four dimensional expressions, induced from geometry in $5D$ STM
theories, in terms of the extra terms appearing in $f(R)$ theories
of gravity. The method is demonstrated by providing an explicit
example in which a given $f(R)$ is used to predict the properties of
the corresponding $5D$ Ricci-flat universe. The accelerated
expansion and the induced dark energy in a 5D Ricci-flat universe
characterized by a big bounce is studied and it is shown that an
arbitrary function $\mu(t)$ in the 5D solutions can be rewritten, in
terms of the redshift $z$, as a new arbitrary function $F(z)$ which
corresponds to the 4D curvature quintessence models.
\end{abstract}
\vspace{2cm}

%%%%%%%%%%%%%%%%%%%%%%%%%%%%%%%%%%%%%%%%%%%%%%%%%%%%%%%%%%%%%%%%%%%%%%%%%%%%%%%
\section{Introduction}

The recent distance measurements from the light-curves of several
hundred type Ia supernovae \cite{1,2} and independently from
observations of the cosmic microwave background (CMB) by the WMAP
satellite \cite{3} and other CMB experiments \cite{4,5} suggest
strongly that our universe is currently undergoing a period of
acceleration. This accelerating expansion is generally believed to
be driven by an energy source called dark energy. The question of
dark energy and the accelerating universe has been therefore the
focus of a large amount of activities in recent years. Dark energy
and the accelerating universe have been discussed extensively from
various point of views over the past few years
\cite{Quintessence,Phantom,K-essence}. In principle, a natural
candidate for dark energy could be a small positive cosmological
constant. One approach in this direction is to employ what is known
as modified gravity where an arbitrary function of the Ricci scalar
is added to the Einstein-Hilbert action. It has been shown that such
a modification may account for the late time acceleration and the
initial inflationary period in the evolution of the universe
\cite{modifidgravity,modifidgravity2}. Alternative approaches have
also been pursued, a few example of which can be found in
\cite{sahni,cardassian,chaplygin}. These schemes aim to improve the
quintessence approach overcoming the problem of scalar field
potential, generating a dynamical source for dark energy as an
intrinsic feature. The goal would be to obtain a comprehensive model
capable of linking the picture of the early universe to the one
observed today, that is, a model derived from some effective theory
of quantum gravity which, through an inflationary period would
result in today accelerated Friedmann expansion driven by some
$\Omega_{\Lambda}$-term. However, the mechanism responsible for this
acceleration is not well understood and many authors introduce a
mysterious cosmic fluid, the so called dark energy, to explain this
effect \cite{Carroll}. As was mentioned above, it has been shown
that such an accelerated expansion could be the result of a
modification to the Einstein-Hilbert action \cite{Deffayet}. A
scenario where the issue of cosmic acceleration in the framework of
higher order theories of gravity in $4D$ is addressed can be found
in \cite{capo}. One of the first proposals in this regard was
suggested in \cite{modifidgravity} where a term of the form $R^{-1}$
was added to the usual Einstein-Hilbert action. In $f(R)$ gravity,
Einstein equations posses extra terms induced from geometry which,
when moved to the right hand side, may be interpreted as a matter
source represented by the energy-momentum tensor $T^{\rm Curv}$, see
equation (\ref{eq03}).

In a similar fashion, the Space-Time-Matter (STM) theory, discussed
below, results in Einstein equations in $4D$ with some extra
geometrical terms which may be interpret as induced matter. It
therefore seems plausible to make a correspondence between the
geometrical terms in STM and $T^{\rm Curv}$ resulting in $f(R)$
gravity. We shall explore this idea to show that different choices
of the parameter $\mu(t)$ in STM may correspond to different choices
of $f(R)$ in curvature quintessence models in modified gravity.

The correspondence discussed above is based on the idea of extra
dimensions. The idea that our world may have more than four
dimensions is due to Kaluza \cite{Kaluza}, who unified Einstein's
theory of General Relativity with Maxwell's theory of
Electromagnetism in a $5D$ manifold. Since then, higher dimensional
or Kaluza-Klein theories of gravity have been studied extensively
\cite{Klein} from different angles. Notable amongst them is the STM
theory mentioned above, proposed by Wesson and his collaborators,
which is designed to explain the origin of matter in terms of the
geometry of the bulk space in which our $4D$ world is embedded, for
reviews see \cite{Wesson}. More precisely, in STM theory, our world
is a hypersurface embedded in a five-dimensional Ricci-flat
($R_{AB}=0$) manifold where all the matter in our world can be
thought of as being  manifestations of the geometrical properties of
the higher dimensional space. The fact that such an embedding can be
done is supported by Campbell's theorem \cite{Compbell} which states
that any analytical solution of the Einstein field equations in $N$
dimensions can be locally embedded in a Ricci-flat manifold in
$\left(N+1\right)$ dimensions. Since the matter is induced from the
extra dimension, this theory is also called the induced matter
theory. Applications of the idea of induced matter or induced
geometry can also be found in other situations \cite{FSS}. The STM
theory allows for the metric components to be dependent on the extra
dimension and does not require the extra dimension to be compact.
The sort of cosmologies stemming from STM theory is studied in
\cite{LiuW,STM-cosmology,WLX}.

In this paper we consider the correspondence between $f(R)$ gravity
and STM theory. In section 2 we present a short review of $4D$ dark
energy models in the framework of $f(R)$ gravity. In section 3 the
field equations are solved in STM theory by fixing a suitable metric
and the resulting geometric terms are interpreted as dark energy.
The cosmological evolution in STM are considered in section 4.
Section 5 deals with an example for a special form of $f(R)$.
Conclusions are drawn in the last section.

%%%%%%%%%%%%%%%%%%%%%%%%%%%%%%%%%%%%%%%%%%%%%%%%%%%%%%%%%%%%%%%%%%%%%%%%%%%%%%%%%%
\section{Modified $f(R)$ gravity}

General coordinate invariance in the gravitational action, without
the assumption of linearity, allows infinitely many additive terms
to the Einstein-Hilbert action \cite{caro}
\begin{equation}\label{eq00}
{\cal S}=\int d^4x \sqrt{-g}[c_0 R + c_1 R^2 + c_2 R_{\mu \nu}R^{\mu
\nu}+ c_3 R_{\mu \nu \lambda \delta}R^{\mu \nu \lambda \delta}+
\cdot\cdot\cdot ]+{\cal S}_{m},
\end{equation}
where $R, R_{\mu \nu}$ and $R_{\mu \nu \lambda \delta}$ are Ricci
scalar, Ricci tensor and Reimann tensor, respectively  and ${\cal
S}_{m}$ is the action for the matter fields. The fourth order term
$R_{\mu \nu \lambda \delta}R^{\mu \nu \lambda \delta}$ may be
neglected as a consequence of the Gauss-Bonnet theorem. The action
(\ref{eq00}) is not canonical because the Lagrangian function
contains derivatives of the canonical variables of order higher than
one. This means that, not only do we expect higher order field
equations, but also the validity of the Euler-Lagrange equations is
compromised. This problem is particularly difficult in the general
case, but can be solved for specific metrics. In homogeneous and
isotropic spacetimes, the Lagrangian in (\ref{eq00}) can be further
simplified. Specifically, the variation of the term $R_{\mu
\nu}R^{\mu \nu}$ can always be rewritten in terms of the variation
of $R^2$. Thus, the "effective" fourth order Lagrangian in cosmology
contains only powers of R and we can suppose, without loss of
generality, that the general form for a non-linear Lagrangian is
given by
\begin{equation}\label{eq01}
{\cal S}=\int d^4x \sqrt{-g} f(R)+{\cal S}_{m},
\end{equation}
where $f(R)$ is a generic function of the Ricci scalar\footnote{We
use units such that $8\pi G_N=c=\hbar=1$.}. Variation with respect
to the metric $g_{\mu \nu}$ leads to the field equations
\begin{equation}\label{eq02}
f'(R)R_{\mu\nu}-\frac{1}{2}f(R)g_{\mu\nu}=
f'(R)^{;\alpha\beta}(g_{\alpha\mu}g_{\beta\nu}-g_{\alpha\beta}g_{\mu\nu})+
\tilde{T}^{m}_{\mu\nu}\,,
\end{equation}
where
\begin{equation}
\tilde{T}^{m}_{\mu \nu}=\frac{2}{\sqrt{-{g}}}\frac{\delta {\cal
S}_{m}}{\delta {g}^{\mu \nu}}
\end{equation}
and the prime denotes a derivative with respect to $R$. It is easy
to check that standard Einstein equations are immediately recovered
if $f(R) = R$. When $f'(R)\neq 0$ the equation (\ref{eq02}) can be
recast in the more expressive form
\begin{equation}\label{eq03}
G_{\mu\nu}=R_{\mu\nu}-\frac{1}{2}g_{\mu\nu}R=T^{\rm
Curv}_{\mu\nu}+T^{m}_{\mu\nu}\,,
\end{equation}
where an stress-energy tensor has been defined for the curvature
contribution
\begin{equation}\label{eq04}
 T^{\rm^{Curv}}_{\mu\nu}=\frac{1}{f'(R)}\left\{\frac{1}{2}g_{\mu\nu}\left[f(R)-Rf'(R)\right]+
f'(R)^{;\alpha\beta}(g_{\alpha\mu}g_{\beta\nu}-g_{\alpha\beta}g_{\mu\nu})
\right\},
\end{equation}
and
\begin{equation}\label{eq05}
 T^{m}_{\mu\nu}=\frac{1}{f'(R)}\tilde{T}^{m}_{\mu\nu}\,,
\end{equation}
is an effective stress-energy tensor for standard matter. This step
is conceptually very important since a gravity model with a
complicated structure converts to a model in which the gravitational
field has the standard GR form with a source made up of two fluids:
perfect fluid matter and an effective fluid (curvature fluid) that
represents the non-Einsteinian part of the gravitational
interaction.

We now consider the Robertson-Walker metric for the evolution of the
universe
\begin{equation}\label{eq06}
ds^{2}=dt^{2}-a(t)^{2}\left(\frac{dr^{2}}{1-kr^{2}}+r^{2}d\Omega^{2}\right),
\end{equation}
where $k$ is the curvature of the space, namely, $k=0, 1, -1$ for
the flat, closed and open universes respectively. Substituting the
above metric with $k=0$ in equation (\ref{eq03}) we obtain the $4D$,
spatially flat Friedmann equations as follows
\begin{equation}\label{eq07}
H^2=\frac{1}{3}\left(\rho_{_{m}}+\rho_{_{\rm Curv}}\right),
\end{equation}
and
\begin{equation}\label{eq08}
\dot{H}=-\frac{1}{2}\left[(\rho_{_{m}}+p_{_{m}})+\rho_{_{\rm
Curv}}+p_{_{\rm Curv}}\right],
\end{equation}
where a dot represents derivation with respect to time. Such a
universe is dominated by a barotropic perfect fluid with the
equation of state (EOS) given by $ p_{_{m}}=w_{_{m}}\rho_{_{m}}$
($w_{m}=0$ for pressureless cold dark matter and $w_{m}=1/3$ for
radiation) and a spatially homogenous curvature quintessence.

The energy density and pressure of the curvature quintessence are
\begin{equation}\label{eq09}
p_{_{\rm
Curv}}=\frac{1}{f'(R)}\left\{2\left(\frac{\dot{a}}{a}\right)\dot{R}f''(R)+\ddot{R}f''(R)+\dot{R}^2f'''(R)
-\frac{1}{2}\left[f(R)-Rf'(R)\right] \right\},
\end{equation}
and
\begin{equation}\label{eq10}
\rho_{_{\rm
Curv}}=\frac{1}{f'(R)}\left\{\frac{1}{2}\left[f(R)-Rf'(R)\right]
-3\left(\frac{\dot{a}}{a}\right)\dot{R}f''(R) \right\},
\end{equation}
respectively. The equation of state of the curvature quintessence is
\begin{equation}\label{eq11}
w_{_{\rm Curv}}=\frac{p_{_{\rm Curv}}}{\rho_{\rm
_{Curv}}}.\label{eosscalar}
\end{equation}
Recently, cosmological observations  have indicated that our
universe is undergoing an accelerated expanding phase. This could be
due to the vacuum energy or dark energy which dominates our universe
against other forms of matter such as dark matter and Baryonic
matter. We thus concentrate on the vacuum sector {\it i.e.}
$\rho_{m}=p_{m}=0$, from which the evolution equation of curvature
quintessence becomes
\begin{equation}\label{eq12}
\dot{\rho}_{_{\rm Curv}}+3H\left(\rho_{_{\rm Curv}}+p_{_{\rm
Curv}}\right)=0,
\end{equation}
which yields
\begin{eqnarray}\label{eq13}
\rho_{_{\rm Curv}}(z)&=&\rho_{0_{\rm Curv}
}\exp\left[3\int_{0}^{z}(1+w_{_{\rm Curv}})d\ln(1+z)\right]\nonumber\\
&\equiv&\rho_{0_{\rm Curv} }E(z),
\end{eqnarray}
where, $1+z=\frac{a_0}{a}$ is the redshift and the subscript $0$
denotes the current value.  In terms of the redshift, the first
Friedmann equation can be written as
\begin{equation}\label{eq14}
H(z)^2=H_0^2\Omega_{0_{\rm Curv}}E(z),\label{Hz}
\end{equation}
where $\Omega_{0_{\rm Curv}}$ and $H_0$ are the current values of
the dimensionless density parameter and Hubble parameter,
respectively. Equation (\ref{eq14}) is the Friedmann equation in
terms of redshift, $z$, which is suitable for cosmological
observations. In fact, equations (\ref{eq14}) and (\ref{eq26}),
obtained in section 4, are the cosmological connections between
$f(R)$ gravity  and STM theory.

%%%%%%%%%%%%%%%%%%%%%%%%%%%%%%%%%%%%%%%%%%%%%%%%%%%%%%%%%%%%%%%%%%%%%%%%%%
\section{Space-Time-Matter theory}\label{DE5}
According to the old suggestion of Kaluza and Klein the $5D$ vacuum
Kaluza-Klein equations can be reduced under certain conditions to
the $4D$ vacuum Einstein equations plus the $4D$ Maxwell equations.
Recently, the idea that our four-dimensional universe might have
emerged from a higher dimensional space–time is receiving much
attention \cite{Freund}. One current interest is to find out in a
more general way how the $5D$ field equations relate to the $4D$
ones. In this regard, a proposal was made recently by Wesson
\cite{Wesson} in that the $5D$ Einstein equations without sources
$R_{AB} = 0$ (the Ricci flat assumption) may be reduced to the $4D$
ones with sources $G_{ab} = 8\pi G T_{ab}$ , provided an appropriate
definition is made for the energy-momentum tensor of matter in terms
of the extra part of the geometry. Physically, the picture behind
this interpretation is that curvature in (4 + 1) space induces
effective properties of matter in (3 + 1) space–time. This idea is
known as {\it space time matter} (STM) or modern Kaluza–Klein
theory.

In this popular non-compact approach to Kaluza-Klein gravity, the
gravitational field is unified with its source through a new type of
$5D$ manifold in which space and time are augmented by an extra
non-compact dimension which induces $4D$ matter within four
dimensional universe. Unlike the usual Kaluza-Klein theory in which
a cyclic symmetry associated with the extra dimension is assumed,
the new approach removes the cyclic condition and derivatives of the
metric with respect to the extra coordinate are retained. This
induces non-trivial matter on the hypersurface of $l = constant$.
This theory basically is guaranteed by an old theorem of
differential geometry due to Campbell \cite{Compbell}.

In the context of STM theory, a class of exact $5D$ cosmological
solutions has been investigated and discussed in \cite{LiuM}. This
solution was further pursued in \cite{LiuW} where it was shown to
describe a cosmological model with a big bounce as opposed to the
ubiquitous big bang. The $5D$ metric of this solution reads
\begin{equation}\label{eq15}
dS^{2}=B^{2}dt^{2}-A^{2}\left( \frac{dr^{2}}{1-kr^{2}}+r^{2}d\Omega
^{2}\right) -dy^{2}, \label{5-metric}
\end{equation}
where $d\Omega ^{2}\equiv \left( d\theta ^{2}+\sin ^{2}\theta d\phi
^{2}\right)$ and
\begin{eqnarray}\label{eq16}
A^{2}&=&\left( \mu ^{2}+k\right) y^{2}+2\nu y+\frac{\nu ^{2}+K}{\mu
^{2}+k},
\nonumber \\
B&=&\frac{1}{\mu }\frac{\partial A}{\partial t}\equiv
\frac{\dot{A}}{\mu }.
\end{eqnarray}
Here $\mu =\mu (t)$ and $\nu =\nu (t)$ are two arbitrary functions
of $t$, $k$ is the $3D$ curvature index $\left(k=\pm 1,0\right)$,
and $K$ is a constant. This solution satisfies the $5D$ vacuum
equation $R_{AB}=0$. The Kretschmann curvature scalar
\begin{eqnarray}\label{eq17}
I_{3} &=&R_{ABCD}R^{ABCD}=\frac{72K^{2}}{A^{8}},
\end{eqnarray}
shows that $K$ determines the curvature of the $5D$ manifold. Such a
solution was considered in \cite{LiuM} with a different notation.

Using the $4D$ part of the $5D$ metric (\ref{eq15}) to calculate the
$4D$ Einstein tensor, we obtain
\begin{eqnarray}\label{eq18}
^{(4)}G_{0}\,^{0} &=&\frac{3\left( \mu ^{2}+k\right) }{A^{2}},
\nonumber \\
^{(4)}G_{1}\,^{1} &=&^{(4)}G_{2}\,^{2}=^{(4)}G_{3}\,^{3}=\frac{2\mu \dot{\mu}}{A%
\dot{A}}+\frac{\mu ^{2}+k}{A^{2}}.
\end{eqnarray}
As was mentioned earlier, since the recent observations show that
the universe is currently going through an accelerated expanding
phase, we assume that the induced matter contains only dark energy
with $\rho_{_{DE}}$, {\it i.e.} $\rho_{_{m}}=0$. We then have
\begin{equation}\label{eq19}
\frac{3\left( \mu ^{2}+k\right) }{A^{2}}=\rho_{_{DE}},
\end{equation}
\begin{equation}\label{eq20}
\frac{2\mu \dot{\mu}}{A\dot{A}}+\frac{\mu ^{2}+k}{A^{2}}=-p_{_{DE}}.
\end{equation}
From equations (\ref{eq19}) and (\ref{eq20}), one obtains the EOS of
dark energy
\begin{equation}\label{eq21}
w_{_{DE}}=\frac{p_{_{DE}}}{\rho_{_{DE}}}=-\frac{2\left. \mu
\dot{\mu}\right/ A \dot{A}+\left. \left( \mu ^{2}+k\right) \right/
A^{2}}{3\left. \left( \mu ^{2}+k\right) \right/ A^{2}}.
\end{equation}
The Hubble and deceleration parameters are given in \cite{LiuW,WLX}
and can be written as
\begin{eqnarray}\label{eq22}
H&\equiv&\frac{\dot{A}}{A B}=\frac{\mu}{A},
\end{eqnarray}
and
\begin{eqnarray}\label{eq23}
q \left(t, y\right)&\equiv&\left.
-A\frac{d^{2}A}{dt^{2}}\right/\left(\frac{dA}{dt}\right)^{2}
=-\frac{A \dot{\mu}}{\mu \dot{A}},
\end{eqnarray}
from which we see that $\dot{\mu}\left/\mu\right.>0$ represents an
accelerating universe while $\dot{\mu}\left/\mu\right.<0$ represents
a decelerating one. The function $\mu(t)$ therefore plays a crucial
role in defining the properties of the universe at late times.

%%%%%%%%%%%%%%%%%%%%%%%%%%%%%%%%%%%%%%%%%%%%%%%%%%%%%%%%%%%%%%%%%%%%%%%%%%%%%%%%
\section{ Correspondence between modified $f(R)$ gravity and STM theory}
In this section we will concentrate on the predictions of the
cosmological evolution in the spatially flat case ($k=0$). To avoid
having to specify the form of the function $\nu(t)$, we change the
parameter $t$ to $z$ and use $A_{0}\left/A \right.=1+z$ and define
$\mu_{0}^{2}\left/ \mu^{2}\right.=F\left(z\right)$, noting that
$F(0)\equiv 1$. We then find that equations
(\ref{eq21})-(\ref{eq23}) reduce to
\begin{eqnarray}\label{eq24}
w_{_{DE}}(z) &=&-\frac{1+\left(1+z\right)d\ln
F\left(z\right)\left/dz\right.}{3},
\end{eqnarray}
and
\begin{eqnarray}\label{eq25}
q_{_{DE}}(z)&=&\frac{1+3\Omega_{_{DE}}w_{_{DE}}}{2}=-\frac{\left(1+z\right)}{2}\frac{d\ln
F\left(z\right)}{dz}.
\end{eqnarray}
There is an arbitrary function $\mu(t)$ in the present $5D$ model.
Different choices of $\mu(t)$ may correspond to different choices of
$f(R)$ in curvature quintessence models in modified gravity. Various
choices of $\mu(t)$ correspond to the choices of $F(z)$. This
enables us to look for the desired properties of the universe via
equations (\ref{eq24}) and (\ref{eq25}). Using these definitions,
the Friedmann equation becomes
\begin{equation}\label{eq26}
H^2=H_0^2(1+z)^2F(z)^{-1}. \label{friedmann}
\end{equation}
This would allow us to use the supernovae observational data to
constrain the parameters contained in the model or the function
$F(z)$. By comparing equation (\ref{eq26}) with equation
(\ref{eq14}), we find that there exists a correspondence between the
functions $f(R)$ and $F(z)$. We thus take $F(z)$ as
\begin{equation}\label{eq27}
F(z)=(1+z)^2\left[\Omega_{0_{\rm Curv}}E(z)\right]^{-1}.\label{fz}
\end{equation}
According to (\ref{eq13}), it is easy to see that the function
$E(z)$ is determined by the particular choice for $f(R)$ which, in
turn, determines the function $F(z)$ through equation (\ref{eq27}).
The evolution of the density components and the EOS of dark energy
may now be derived. To this end, we must determine the functional
form of $f(R)$. Thus, for example, we choose $f(R)$ as a generic
power law of the scalar curvature and assume for the scale factor a
power law solution in $4D$, investigated in \cite{capo}. Therefore
\begin{equation}\label{eq28}
f(R)=f_0 R^n\,,\qquad a(t)=a_0\left(\frac{t}{t_0}\right)^{\alpha}\,.
\end{equation}
The interesting cases are for the values of $\alpha$ satisfying
$\alpha> 1$ which would lead to an accelerated expansion of our
universe. Let us now concentrate on the case $\rho_{_{m}}=0$.
Inserting equation (\ref{eq28}) into the dynamical system
(\ref{eq07}) and (\ref{eq08}), for a spatially flat space-time we
obtain an algebraic system for parameters $n$ and $\alpha$
 \begin{equation} \label{eq29}
 \left\{ \begin{array}{ll} \alpha \left[\alpha(n-2)+2n^{2}-3n+1\right]=0, \\
 \\
\alpha\left[n^{2}-n+1+\alpha(n-2)\right]=n(n-1)(2n-1),\\
\end{array}
\right.
\end{equation}

\vspace{7mm}

\noindent from which the allowed solutions are
\begin{equation}\label{eq30}
\begin{array}{cc} \alpha=0\,\, \rightarrow\,\, n=0,\,\,1/2,\,\,1,\\ \\
\alpha=\displaystyle\frac{2n^2-3n+1}{2-n}\,,\,\, \forall{n},\,\ \,
n\neq {2}\,.
\end{array}
\end{equation}
\vspace{7mm} The solutions with $\alpha=0$ are not interesting since
they provide static cosmologies with a non-evolving scale factor.
Note that this result matches the standard General Relativity result
$n=1$ in the absence of matter. On the other hand, the cases with
generic $\alpha$ and $n$ furnish an entire family of significant
cosmological models. Using equations (\ref{eq09}) and (\ref{eq10})
we can also deduce the equation of state for the family of solutions
$\alpha=\displaystyle\frac{2n^2-3n+1}{2-n}$ as \vspace{7mm}
\begin{equation}\label{eq31}
w_{_{\rm Curv}}(n)=-\left(\frac{6n^2-7n-1}{6n^2-9n+3}\right)\,,
\end{equation}
where $w_{_{\rm Curv}}\rightarrow{-1}$ as $n\rightarrow {\infty}$.
This shows that an {\it infinite} $n$ is compatible with recovering
an {\it infinite} cosmological constant. Thus, using equation
(\ref{eq31}), $E(z)$ and $F(z)$ are given by
\begin{eqnarray}\label{eq32}
\label{pot1} E(z)&=&(1+z)^{3\left[\frac{-2n+4}{6n^2-9n+3}\right]},
\end{eqnarray}
\begin{eqnarray}\label{eq33}
\label{pot1} F(z)&=&(1+z)^2\left[\Omega_{0_{\rm Curv}
}(1+z)^{3\left[\frac{-2n+4}{6n^2-9n+3}\right]}\right]^{-1}.
\end{eqnarray}
Now, using the above equations, equations (\ref{eq24}) and
(\ref{eq25}) can be written as
\begin{equation}\label{eq34}
w_{_{DE}}(n)=-\left(\frac{6n^2-7n-1}{6n^2-9n+3}\right),
\end{equation}
and
\begin{equation}\label{eq35}
q_{_{DE}}(n)\equiv-\frac{A\dot{\mu}}{\mu\dot{A}}=\frac{-2n^{2}+2n+1}{2n^{2}-3n+1}.
\end{equation}
Therefore, within the context of the present investigation, the
accelerating, dark energy dominated universe, can be obtained by
using the correspondence between $F(z)$ and $f(R)$ in modified
gravity theories. We observe that in STM theory, $5D$ dark energy
cosmological models correspond to $4D$ curvature quintessence
models. This result is consistent with the correspondence between
exact solutions in Kaluza-Klein gravity and scalar tensor theory
\cite{billyard}. Note that, as is well known, with a suitable
conformal transformation, $f(R)$ gravity reduces to the scalar
tensor theory.

From equations (\ref{eq32}) and (\ref{eq33}), we can rewrite
equation (\ref{eq26}) as
\begin{equation}\label{eq36}
h(z, n)=\Omega_{0_{\rm
Curv}}(1+z)^{3\left[\frac{-2n+4}{6n^2-9n+3}\right]},
\end{equation}
where $h(z,n)\equiv\frac{H(z)^{2}}{H_0^{2}}$ and the contribution of
ordinary matter has been neglected. Figure 1 shows the behavior of
$h(n)$ as a function of $n$ for $z\sim1.5$ and $\Omega_{0_{\rm
Curv}}\simeq0.70$. As can be seen, for $n\longrightarrow\pm\infty$
and $z\longrightarrow0$ we have $h(z,n)\longrightarrow\Omega_{0_{\rm
Curv}}$, that is, the universe finally approaches the curvature
dominant state, thus undergoing an accelerated expanding phase.
Figure 2 shows the behavior of $h(z)$ as a function of $z$ for
$n=2,10,-10$ and $\Omega_{0_{\rm Curv}}\simeq0.70$. We see that for
small $z$, $h(z)\longrightarrow0.70$. Thus, we have obtained
late-time accelerating solutions only by using the correspondence
between $f(R)$ gravity and STM theory. Here, we have interpreted the
properties of $5D$ Ricci-flat cosmologies by dark energy models in
modified gravity.
\begin{figure}
\begin{center}
\epsfig{figure=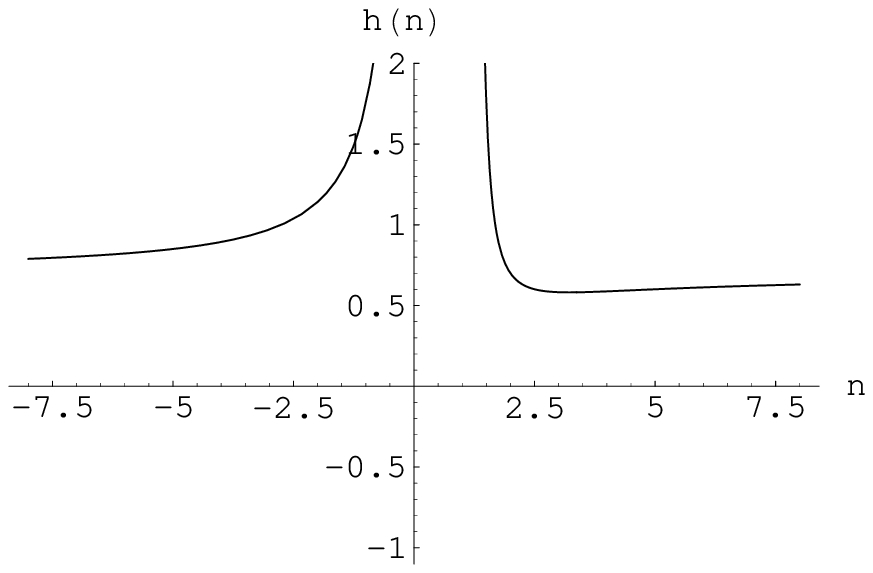,width=9cm}\hspace{5mm}
\end{center}
\caption{\footnotesize Behavior of $h(n)$ as a function of $n$ for
$z\sim1.5$ and $\Omega_{0_{\rm Curv}}\simeq0.70$. An accelerating
universe occurs for $n\lesssim-2$ and $n \gtrsim 2$.}
\begin{center}
\epsfig{figure=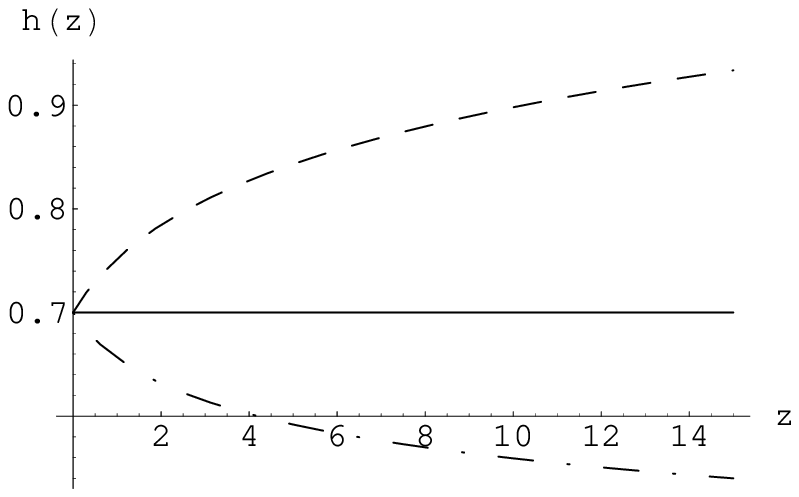,width=7cm}\hspace{5mm}
\end{center}
\caption{\footnotesize Behavior of $h(z)$ as a function of $z$ for
$n=2$ (solid line), $n=10$ (dashed line), $n=-10$ (dot-dashed line)
and $\Omega_{0_{\rm Curv}}\simeq0.70$. Note that for $n=2, 10,-10 $,
$z\rightarrow0$ and $h(z)\rightarrow0.7$ .}
\end{figure}

%%%%%%%%%%%%%%%%%%%%%%%%%%%%%%%%%%%%%%%%%%%%%%%%%%%%%%%%%%%%%%%%%%%%%%%%%%%%%%%%%%%%%
\section{Conclusions}
In this paper we have studied the correspondence between modified
$f(R)$ gravity and Space-Time-Matter theory by investigation of the
present accelerated expanding phase of the universe using a general
class of $5D$ cosmological models, characterized by a big bounce as
opposed to a big bang, which is the standard prediction in $4D$
cosmological models. Such an exact solution contains two arbitrary
functions, $\mu(t)$ and $\nu(t)$, which are analogous to different
forms of $f(R)$ in curvature quintessence models. Also, once the
forms of the arbitrary functions are specified, the characteristic
parameters determining the evolution of our universe are specified.
We have noted that the correspondence between the functions $F(z)$
and $f(R)$ plays a crucial role and defines the form of the function
$F(z)$. Finally, by taking a specific form for $f(R)$ we obtained
solutions that describe the late-time acceleration of the universe.
Explicitly, the induced dark energy and the resulting accelerated
expansion in a 5D Ricci-flat universe is studied and it is shown
that an arbitrary function $\mu(t)$ in the 5D solutions can be
rewritten as a new arbitrary function $F(z)$ which corresponds to
the 4D curvature quintessence models.
%%%%%%%%%%%%%%%%%%%%%%%%%%%%%%%%%%%%%%%%%%%%%%%%%%%%%%%%%%%%%%%%%%%%%%%%%%%%%%%%%%

\end{document}